\documentclass[twocolumn]{elsart}

\usepackage[dvips]{graphicx}
\usepackage{amssymb}

\begin{document}

\newcommand{\myfigcmd}[3]{
     \begin{center}
     \includegraphics[width=\columnwidth,keepaspectratio=true]{#1.eps}
     \mycaption{#2}{#3}
     \label{f:#1}
     \end{center}
}

\newcommand{\mywidefigcmd}[3]{
     \begin{center}
     \includegraphics[width=\textwidth,keepaspectratio=true]{#1.eps}
     \mycaption{#2}{#3}
     \label{f:#1}
     \end{center}
}

\newcommand{\mycaption}[2]{
 \caption[#1]{\textsf{#1.} \footnotesize{\linespread{1}#2}}
}

\begin{frontmatter}


\title{Performance of Hamamatsu 64-anode
photomultipliers for use with wavelength--shifting optical fibres}


\author[ox]{N. Tagg \thanksref{corresp}}
\author[ox]{A. De Santo \thanksref{holloway}}
\author[ox]{A. Weber}
\author[ox]{A. Cabrera}
\author[ox]{P. S. Miyagawa}
\author[ox]{M. A.  Barker \thanksref{BNFL}}
\author[texas]{K. Lang}
\author[caltech]{D. Michael}
\author[ucl]{R. Saakyan}
\author[ucl]{J. Thomas}

\address[ox]{University of Oxford, Denys Wilkinson Building, Keble
Road, Oxford, OX1~3RH, UK}

\address[texas]{University of Texas, Austin, TX 78712, USA}

\address[caltech]{California Institute of Technology, Pasadena, CA 9125,
USA}

\address[ucl]{University College London, Gower Street, London, WC1E~BT, UK}

\thanks[corresp]{Corresponding author. Email: n.tagg1@physics.ox.ac.uk}
\thanks[holloway]{Current address: Royal Holloway, University of London, Egham,
Surrey, TW20~0EX, UK}
\thanks[BNFL]{Current address: BNFL, Sellafield, Seascale, CA20~1PG, UK}

%

\begin{abstract}
Hamamatsu R5900-00-M64 and R7600-00-M64 photomultiplier tubes will be
used with wavelength--shifting optical fibres to read out scintillator
strips in the MINOS near detector. We report on measurements of the
gain, efficiency, linearity, crosstalk, and dark noise of 232 of these
PMTs, of which 219 met MINOS requirements.
\end{abstract}

\end{frontmatter}

\section{Introduction} \label{s:intro}

MINOS, a long baseline neutrino-oscillation experiment, uses two large
segmented tracking calorimeters to make precise measurements of the
atmospheric neutrino oscillation parameters
\cite{nove,Michael:2003fg}.   MINOS uses extruded scintillator strips
to form the calorimeter. Wavelength-shifting (WLS) fibres are glued into a
longitudinal groove along each strip, so that some of the blue
scintillation light is absorbed in the fibre and isotropically
re-emitted as green light. A fraction of green light is trapped in the
fibre and transmitted along it. At the end of the strip, clear
polystyrene fibres carry the light to multi-anode PMTs. Planes of
1~cm thick scintillator strips are sandwiched between 2.54~cm thick
planes of steel to form the detectors.

In the MINOS Far Detector, R5900-00-M16 PMTs
\cite{Lang:2001tx,companion-paper,Korpar:2000iy} are used with 8 fibres optically
coupled to each pixel in order to reduce the cost of the readout
electronics.  At the Near Detector, MINOS uses R5900-00-M64 and
R7600-00-M64 PMTs\footnote{ The R5900-00-M64 and R7600-00-M64 models
differ only slightly in that the latter lacks an external mounting
flange. Hamamatsu has replaced the R5900 model with the R7600
\cite{private-comm-ham}.} (collectively referred to in this work as ``M64s'')
with one fibre per pixel to avoid reconstruction ambiguities
in the higher-rate detector. 

Several other experiments have needs similar to MINOS and the same
basic technology for reading out scintillator. In particular, OPERA
\cite{Migliozzi:2003gz} will use M64 PMTs in a tracker similar
to MINOS detectors. The proposed MINER$\nu$A experiment \cite{minerva}
has also chosen M64s as their baseline technology. The K2K SciBar
detector uses similar multi-anode PMTs
\cite{Nitta:2004nt}. Scintillating fibre detectors have very similar
requirements for PMTs, and M64s have been proposed or adopted by
HERA-B \cite{hera}, CALET \cite{calet}, GLAST \cite{glast}, PET
detectors \cite{pet}, and neutron detectors \cite{neutron}. M64s are
also been studied for suitability in RICH detectors by LHCb
\cite{lhcb}.

In this work we present the analysis of 232 M64 PMTs (13 of which were
R7600, the rest of which were R5900) from several production batches,
delivered between May 2001 and August 2003.  The PMTs included in this
sample were those that passed extensive testing and review; out of 232
PMTs tested, 13 PMTs were rejected for use in MINOS by criteria
discussed below. The 13 rejected PMTs were not included in the plots
and other results presented here.

The basic description of the M64 and the test equipment is described
in section \ref{s:descrip}.  For use in MINOS, the PMTs were required
to have well-resolved single-photoelectron peak for each pixel, as
described in section \ref{s:1pe}, to ensure high sensitivity. Because
of the limited dynamic range of MINOS electronics, a specified
uniformity was required between pixels on a single PMT.  Uniformity
constraints on both gain and efficiency are defined in section
\ref{s:uni}.  Linearity, described in section \ref{s:lin},  was
measured and was required to reduce the
systematic effects when doing calorimetry on high-density
showers. The multi-channel nature of the device raised concerns about
crosstalk, which is discussed in section
\ref{s:xtalk}. Finally, low dark noise, discussed
in section \ref{s:dark}, is a required feature for MINOS as it reduces
load on the data acquisition system.

\section{Description of the M64 and test equipment} \label{s:descrip}

The M64 PMT consists of a single bi-alkali photocathode behind which
are focusing electrodes which guide photoelectrons into one of 64
pixels arranged on an 8 by 8 grid. Each pixel is multiplied by a two
``metal dynode channels''; each dynode plate has two slits per pixel
which act as the multiplying surfaces. Focusing wires are used to keep
electrons inside the logical pixel areas. Each pixel is read out by a
single anode pad.  The active area of each pixel on the PMT window is
approximately 1.4~mm$\times$1.4~mm. Between pixels is a 0.3~mm space
in which efficiency is reduced.  The R5900-00-M64 PMT is shown in
Figure \ref{f:pmt_pic}.  

\begin{figure}
\myfigcmd{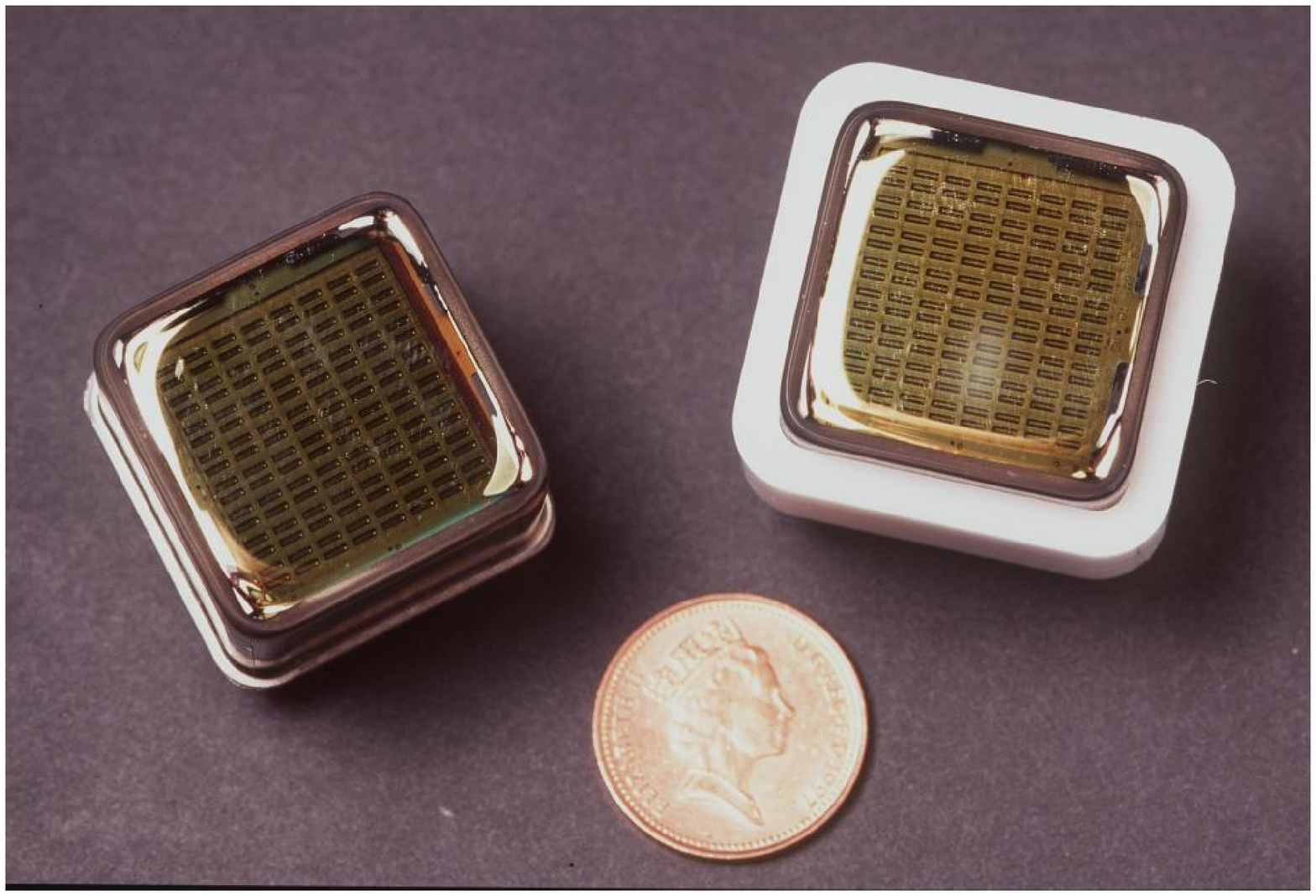}{Picture of the M64}{The M64 on the right is shown
with the fitted Norel plastic collar. The PMT on the left is an
R5900-00-M64, with the small flange near the base of the PMT.}
\end{figure}

The MINOS experiment operates PMTs with the cathodes at negative high
voltage and the anodes at ground. A custom-made printed circuit board
attached to the base of the PMT provides a voltage-divider circuit to
apply voltage to the dynodes in the ratio recommended by
Hamamatsu\cite{datasheet} for optimal performance
(3:2:2:1:1:1:1:1:1:1:1:2:5). Capacitors are used at the last stages to
stabilize the potentials in the case of large instantaneous
currents. In addition to the 64 anode signals, a capacitive tap on the
12th dynode was provided on the PCB. The charge on this dynode signal
was integrated to provide a simple analog sum of the 64 channels. In
the MINOS detector electronics, this dynode signal is used for
triggering readout.


\begin{table*}
\begin{center}
\begin{tabular}{|| l | l ||}
\hline \hline
{\em M64 PMTs} & \\
Window    		& 0.8~mm borosilicate \\ 
Casing			& KOVAR metal \\
Size			& 28 mm x 28 mm x 20 mm \\
Weight			& 28 g \\
Photocathode 		& bi-alkali \\
Dynode type		& metal channel, 12 stages \\
Spectral response 	& 300 to 650 nm \\
Peak Sensitive Wavelength  		& 420 nm\\
Anode dark current	& $\le$ 0.2 nA per pixel \\
Maximum HV		& 1000 V \\
Gain at 800 V		& $\sim 3 \times 10^5$ (typ.) \\
Anode rise time		& 1.5 ns \\
Transit spread time	& 0.3 ns FWHM \\
Pulse linearity		& 0.6 mA per channel \\
Pixel uniformity	& 1:3(max) $\sim$30\%(RMS)\\
\hline
{\em WLS Fibre} & \\
Manufacturer model	&  Kuraray double-clad 1.2~mm diameter \\
Material                &  Polystyrene and polyfluor  \\
Fibre Fluor	        &  Y11 \\
Fluor Decay time       &  $\sim$7 ns \\
\hline
{\em Clear Fibre} & \\
Manufacturer model	&  As WLS, without fluor \\
\hline
{\em Light Source} & \\
Source			& 5~mm ``ultra-bright'' blue LED \\
Pulse width             & $<$ 5 ns \\
Peak wavelength 	& 470 nm \\
\hline \hline
\end{tabular}
\label{t:props}
\mycaption{Properties of the M64 PMTs,Fibres, and light source}{Nominal characteristics of the
PMTs are taken from Hamamatsu, Ref.~\cite{datasheet}}
\end{center}
\end{table*}

Before testing, the M64s were mounted into the MINOS PMT assembly
hardware similar to that described in
Ref.~\cite{companion-paper}. First, M64s were glued into uniform Norel
collars. These collars were designed to slide tightly into a ``PMT
holder''. The voltage-divider PCB was attached to the back of the
holder, and the front of the holder was attached a ``cookie holder'', which
provided attachment points for the optical fibre mount. Because the
shape of the outer PMT casing does not have a fixed relation to the
pixel positions, the cookie holder was aligned with respect to the PMT
such that it was centered on alignment marks etched in the first
dynode plate.  This arrangement allows any fibre cookie to be attached
to any PMT with the fibres in the correct positions centered above the
pixels.  The alignment system had a precision better than
0.1~mm. (Previous measurements \cite{early_paper} have indicated that
this precision is adequate to center the fibres. At this precision, the PMT
response is reproducible after disassembly and re-alignment.)

\begin{figure*}
\mywidefigcmd{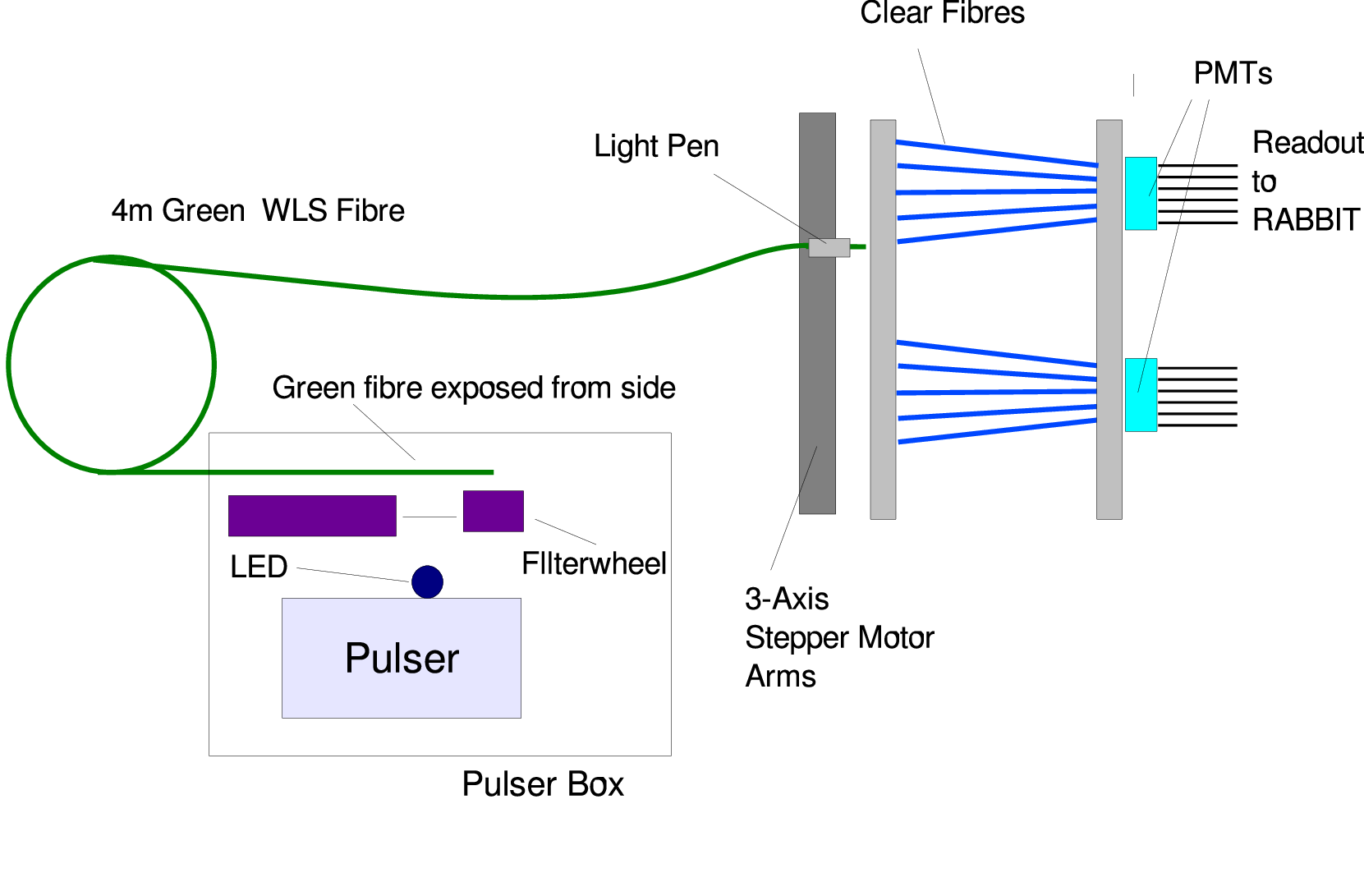}{Test Stand Illustration}{Light from an LED pulser
is flashed through a changeable filter onto the side of a WLS fibre,
which carries the light to one of 64 clear fibres for each of four
PMTs. The polished clear fibres are pressed against the PMT face in
alignment with the pixel positions.}
\end{figure*}

For this work, the PMTs were mounted on a test stand illustrated in
Figure \ref{f:diagram}.  Each of three PMTs had 64 clear fibres routed
to it. The clear fibres were 1.2~mm in diameter, and were fly-cut with
a diamond bit to give a polished finish. The PMT was mounted
with alignment pins that ensured that the clear fibres were centered
on the active pixel areas of the PMT.

Each clear fibre terminated at the center of a hole in
an aluminum plate. A stepper-motor system was constructed to move a
cylinder (the ``light pen'') into any one of these holes. The pen held
the end of a green ($\sim$530~nm) wavelength-shifting fibre.  The
other end of the 4~m WLS fibre\footnote{A four-meter fibre was used to
approximate the attenuated light spectrum seen at the end of a MINOS
scintillator strip.} was illuminated from the side by a blue LED.


The spectrum of green light from the WLS fibre is shown in Figure
\ref{f:spectrum2}\cite{li-paper}. The fibre and LED were
separated by changeable neutral-density filters which provided
different light levels in the tests. The LED was pulsed with a $<$5~ns
wide pulse. The WLS fluor has a decay time of 7~ns, so the light pulse
at the PMT was $\lesssim$10~ns FWHM.

\begin{figure}
\myfigcmd{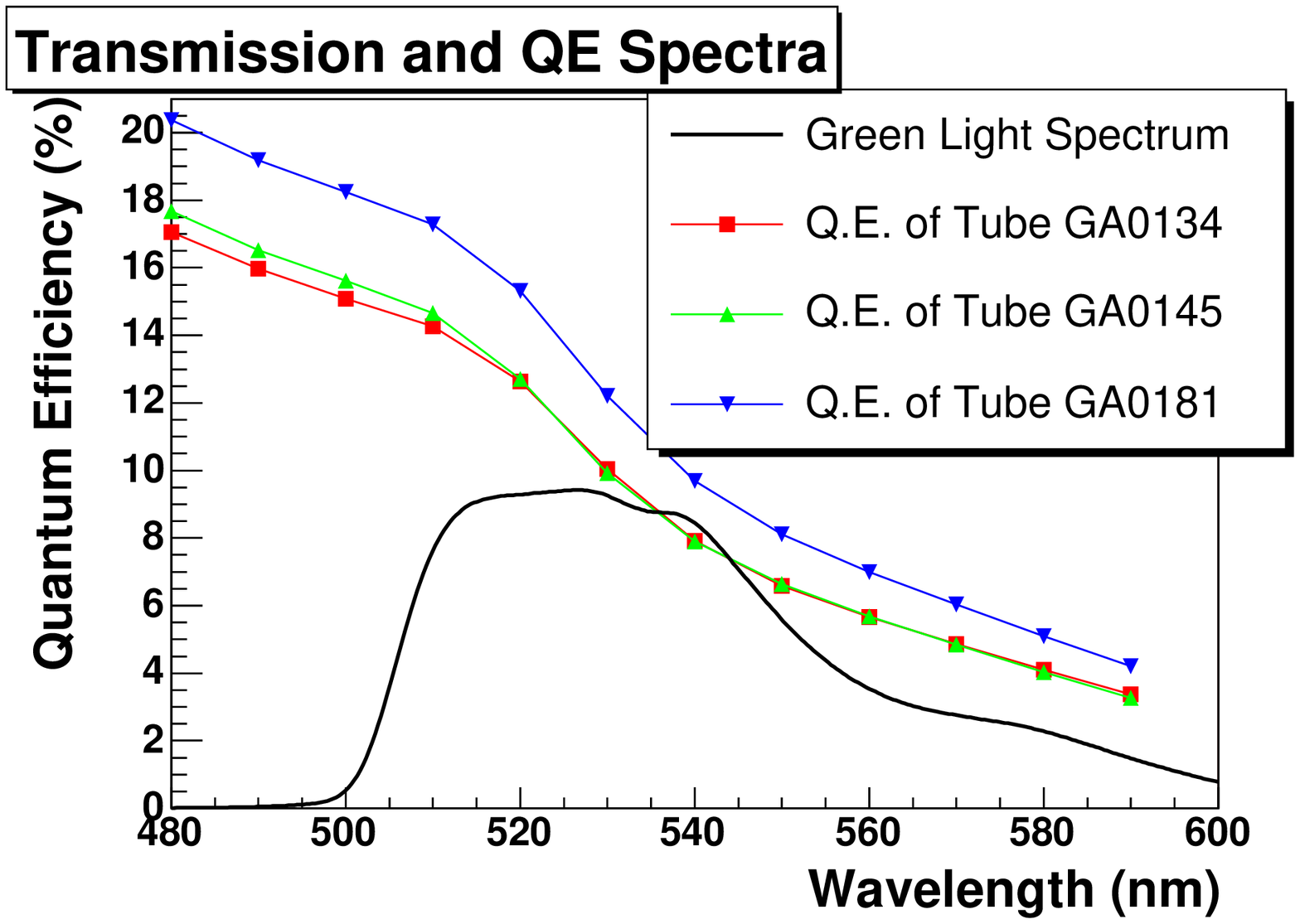}{Light spectrum and quantum efficiency curve}{The
spectrum of light seen at the end of the green fibre\cite{li-paper} is shown (with
arbitrary normalization) relative to the quantum-efficiency curves
provided by Hamamatsu for three M64 PMTs.}
\end{figure}

In these tests, every channel of the PMT was read out with
charge-integrating RABBIT photomultiplier
electronics\cite{rabbit}. The LED was pulsed by the RABBIT system.
The pulse rate was limited by the data acquisition to about 200 pulses
per second. This system used a 16-bit ADC to sequentially read out
each channel with a quantization of 0.71~fC per count and a typical
resolution of 13~fC RMS per channel.  Charges were integrated over a
duration of 1.1~$\mu$s.

Up to three tubes were mounted in the test stand at one time. The
fourth position was permanently occupied by an M64 PMT used for
monitoring the light level. For each LED flash, one pixel on one PMT
was illuminated, and all 64 pixels were read out along with 3 pixels
on the monitoring tube.  For each complete scan of pixels on the PMT,
the three monitor PMT pixels were also flashed to track the light
level of the light pulser. Pedestals were subtracted using data taken
with the PMT unilluminated.  

A complete testing cycle of three PMTs took approximately three
days. Four hours were spent allowing tubes to condition to high
voltage in the dark. Then scans were taken at varying high voltages
for five hours. A nominal operating voltage was chosen automatically,
as described in section \ref{s:uni}. The tubes were then scanned at
each of 11 different light levels (set by the adjustable filter) to
test single-photoelectron response, gain, linearity, and crosstalk over the
next 13 hours. Then illumination was turned off and dark noise was
recorded for the remaining 55 hours, interrupted by two scans to test
the stability of the gain measurements.


\section{Single Photoelectron Response} \label{s:1pe}

Figure \ref{f:single_pe_example} shows the typical charge response for
a PMT illuminated at a light level of approximately 1
photoelectron (p.e.) per pulse, with $10\,000$ pulses per histogram.
The pedestal peak can usually be distinguished from the single--p.e.
curve.  The voltage is set to the ``operating'' high voltage described
in the following section.

The relative RMS width of the single--p.e. peak is approximately 50\% of
the mean charge.  This large fractional width is due in part to the
finite secondary emission ratio of the first dynode, and in part
because the two metal dynode channels in each pixel will in general have slightly
different gains.  In addition to these effects, the electronics
resolution broadens the distribution by an additional $\sim$8\%.

A good fit to the single-p.e. charge spectra was achieved with
Eq. (\ref{eq:fit}). First, the mean number of photoelectrons $\overline{N}_{pe}$
was used to create a Poisson distribution of $n$ photoelectrons
$P(n|\overline{N}_{pe})$.  Then, for each $n$ photoelectrons, the distribution of
$m$ secondary electrons was again chosen as a Poisson distribution
$P(m|n \epsilon)$, where $n \epsilon$ represents the mean number of
secondary electrons.  Each value of $m$ is in turn represented by a
Gaussian of peak position $mge/\epsilon$ and width
$\sqrt{m/\epsilon}(ge/\epsilon)$, where $m$ is the number of secondary
photoelectrons, $g$ is the gain, $e$ is the electron charge, and
$\epsilon$ is a fit parameter describing the width of the single
p.e. peak.  The parameter $\epsilon$ is analogous to the secondary
emission ratio of the first dynode, but cannot be interpreted as such
due to the broadening effect of two dynode channels\cite{numi_note}.

\begin{equation} 
\label{eq:fit}
\begin{array}{ll}
F(q) = & \sum_{n} \frac{(\overline{N}_{pe})^n \exp{(-\overline{N}_{pe})}}{n!} \\
       & \times \sum_{m} \frac{(n \epsilon)^m \exp{(-n \epsilon)}}{m!} \\
       & \times \frac{1}{\sqrt{2 \pi} \sigma}
         \exp\left(- \frac{(q-m g e/{\epsilon})^2}{2 \sigma^2} \right) \\
\mathrm{where} & \sigma = \sqrt{\frac{m}{\epsilon}}\frac{ge}{\epsilon}
\end{array}
\end{equation}

The measured spectra were fit with a Gaussian pedestal peak (with fit
parameters of mean, width, and integral) plus the single-p.e. shape
given by Eq. \ref{eq:fit} (with fit parameters $\overline{N}_{pe}$,
$\epsilon$ and $g$).  

Using the fit values of $g$ and $\epsilon$, the fractional width ($w$) of
the single-p.e. peak was characterized. The average
RMS width of all pixels was 43\%~$\pm$~6\% of the peak position. (This
corresponds to $\epsilon = 5.4 \pm 1.2$.)  The most extreme pixels had
widths as high as 58\% and as low as 35\%.

Five of the 231 PMTs were rejected from the sample for having
poor single-p.e. responses: two had totally dead pixels, and three had one
or more pixels with very wide or indistinguishable single-p.e. charge
spectra.
\begin{figure}
\myfigcmd{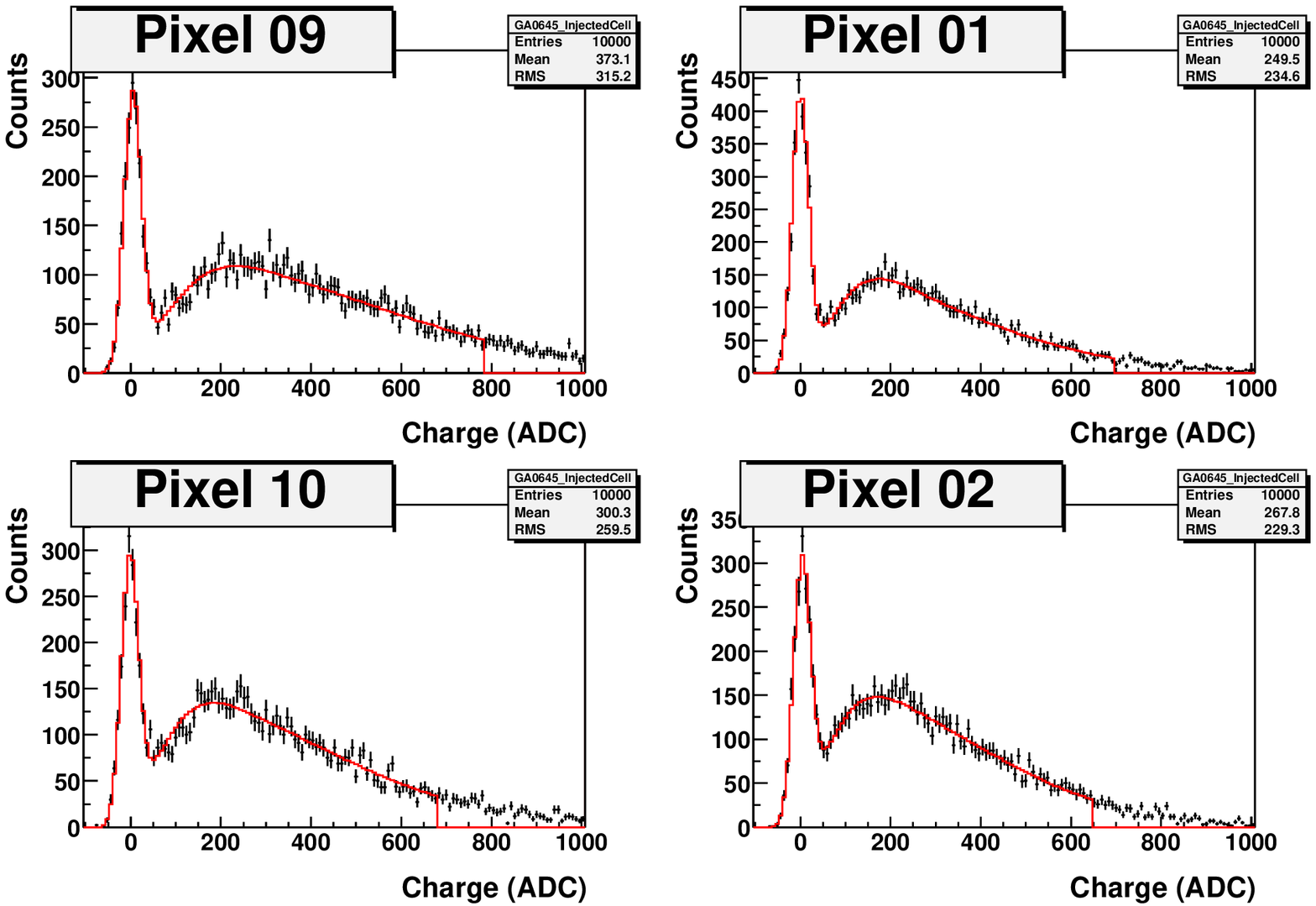}{Example of single-photoelectron
spectra}{Each histogram corresponds to a single pixel on a typical
PMT. The fit to the data is shown by the curved line which goes
through the data points. Units are ADC counts.}
\end{figure}

\section{Gains, Efficiencies, and Pixel Uniformity} \label{s:uni}

To test the pixel gains and efficiencies, 10~000 light injections were
performed on each pixel at a light level of approximately 10 p.e. per
pulse.  These data were used to compute the gain and efficiency of
each pixel, using photon statistics.

We define $\overline{Q}$ as the mean charge of the distribution,
$\sigma_Q$ as the RMS of the charge distribution, and $\sigma_{ped}$ as
the electronics resolution (i.e. the pedestal width), all converted
from ADC counts to units of charge (using the gain of the
RABBIT electronics). The fractional width of the single-p.e. distribution
is given by $w$.  If the mean number of p.e. per pulse is
$\overline{N}_{pe}$, then the gain $g$ is defined as 
\begin{equation}
g = \frac{\overline{Q}}{\overline{N}_{pe} \times e} \label{eq:g}
\end{equation}
where $e$ is the charge of the electron.

The total variance of the charge distribution is the sum of three
terms: the variance of the poisson number of photoelectrons created
each pulse, the variance due to the finite width of the
single-p.e. spectrum, and the variance due to electronic noise:
\begin{eqnarray}
\sigma_Q^2 &=& \left( \sqrt{\overline{N}_{pe}} g e \right)^2
+ \left( \sqrt{\overline{N}_{pe}} g e w\right)^2 \nonumber \\
&&+ \left(\sigma_{ped} \right)^2 \label{eq:var}
\end{eqnarray}

The gain and number of photoelectrons can then be solved, shown in
equations (\ref{eq:npe_corr}) and (\ref{eq:g_corr}), using only the
mean and RMS of the measured charge distribution.

\begin{eqnarray}
\bar{N}_{\rm pe} & = & \frac{\overline{Q}^2}{\sigma_Q^2-\sigma_{ped}^2}
\times \left(1 + w^2 \right) \label{eq:npe_corr} \\
\frac{1}{g} & = & \frac{\overline{Q}}{\sigma_Q^2-\sigma_{ped}^2}
\times \left(1 + w^2 \right) \times e \label{eq:g_corr} 
\end{eqnarray}

The electronics resolution $\sigma_{ped}$ was typically about 13 fC.
For each pixel, the fractional single-p.e. width $w$ was taken to be
50\%; this simplification created a systematic error on $g$ and
$\overline{N}_{pe}$ of only a few percent, similar to the statistical
error. Measurements of the gain and efficiency by this method agreed
well with the results from the single-p.e. fits described in section
\ref{s:1pe}.

The gain of the PMTs were measured at 750, 800, 850, 900, and
950~V. The mean gain of the 64 pixels at each voltage was calculated and
fit to a second-order polynomial for each PMT. One such fit is
shown in Figure \ref{f:gain_vs_hv_example}. An operating voltage was
found for each PMT such that the mean gain was $0.8 \times 10^6$.
The typical slope of the gain curve near the operating HV (expected to
be $\frac{\Delta g}{g} \simeq 12 \frac{\Delta V}{V}$ for 12 dynode
stages) was 1.5\%/V.  The distribution of operating voltages for our
sample is shown in Figure \ref{f:voltages}.  All subsequent
measurements were performed at these operating voltages.

\begin{figure}
\myfigcmd{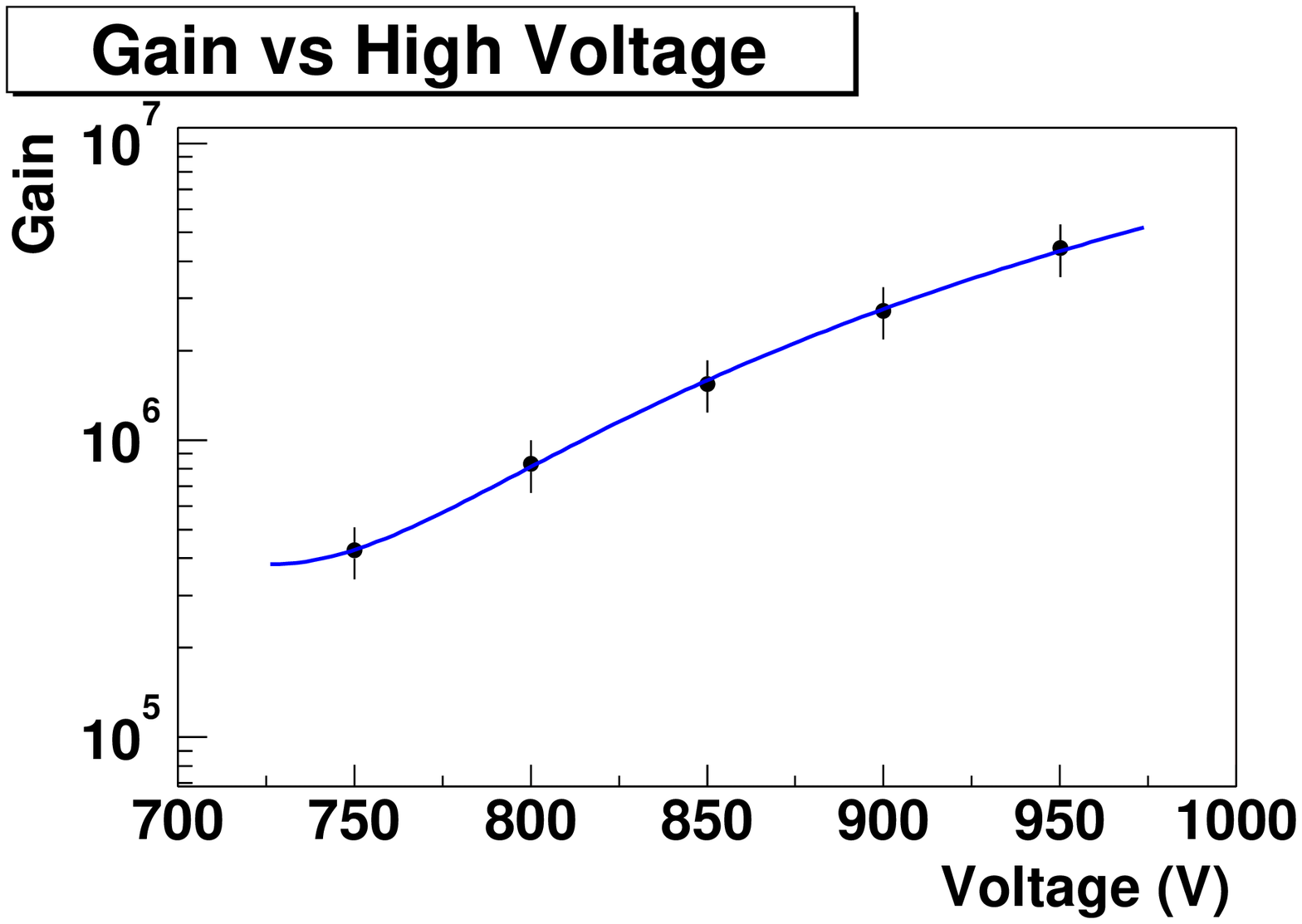}{Example Mean Gain Change With High
Voltage}{The gain change with high voltage is shown for the mean of
all pixels on a typical PMT. The vertical error bars represent the
RMS spread of the individual pixels. An
empirical fit to a simple 2nd order polynomial function is shown by the line.}
\end{figure}

\begin{figure}
\myfigcmd{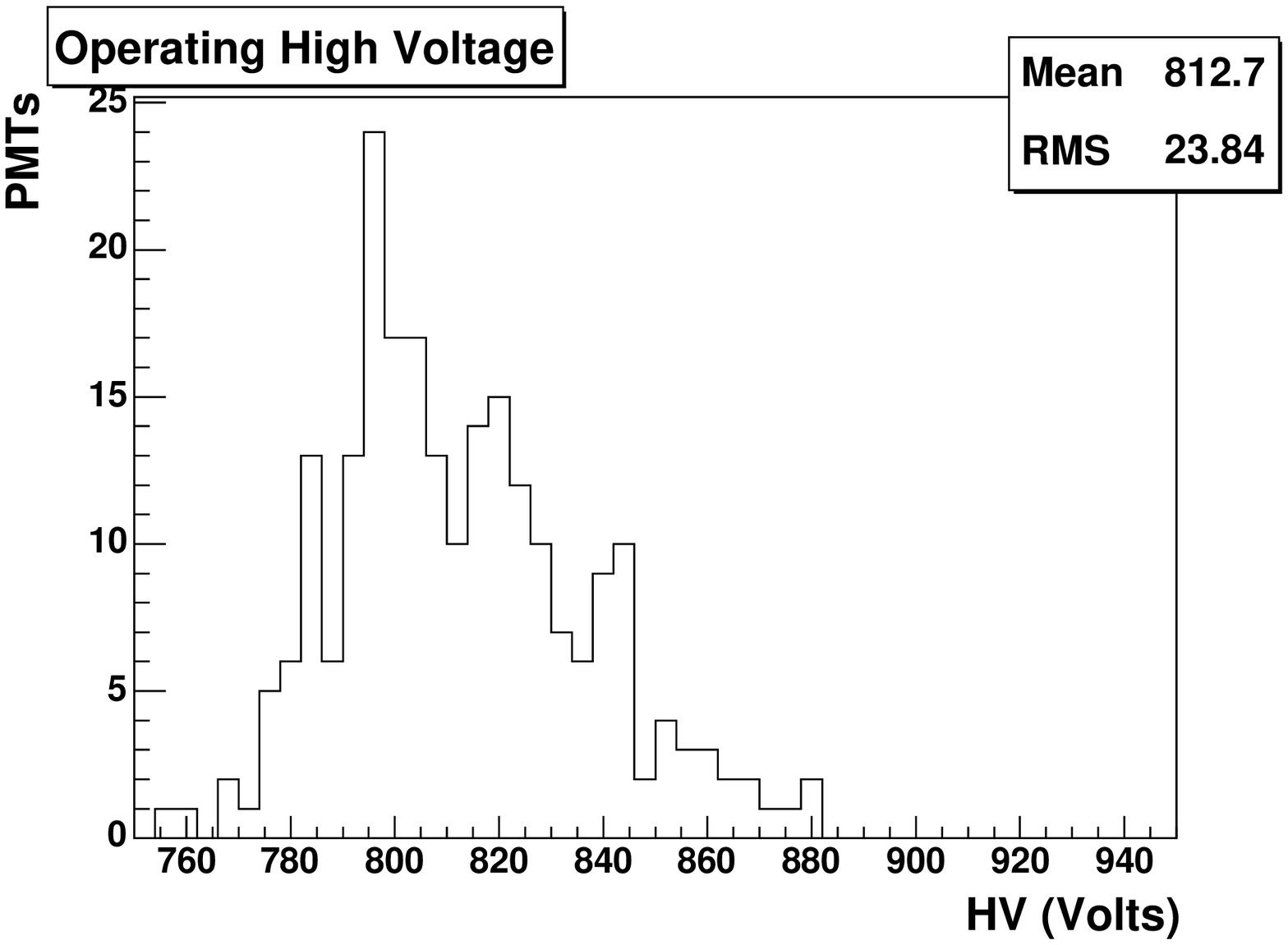}{Operating High Voltage}{The histogram shows the 
voltages necessary to achieve a mean gain per pixel of $0.8 \times 10^6$.}
\end{figure}

The uniformity of the pixel gains on a single PMT was within 15 to
25\% RMS for all accepted PMTs.   The histogram of all the
pixel gains is shown in Figure \ref{f:pixel_gain}. Accepted PMTs were required to have
a maximum-to-minimum gain ratio of less than about 3 to 1.  The histogram of this
ratio is shown in Figure \ref{f:gainMaxMin}.

The gain of the pixels followed a repeatable pattern on most PMTs,
shown in Figure \ref{f:mean_pixel_gain}. In particular, pixels 1--8
and 57--64 tended to show the lowest gains on the PMT. (These pixels
had larger sensitive photocathode area than the others, so the charge
response of the pixels may be more uniform in applications other than
fibre readout.) The high--gain pixels near the bottom of the figure
are near a small vent used to evaporate the photocathode material
during manufacturing; it is possible that this was related to the gain
pattern.

\begin{figure}
\myfigcmd{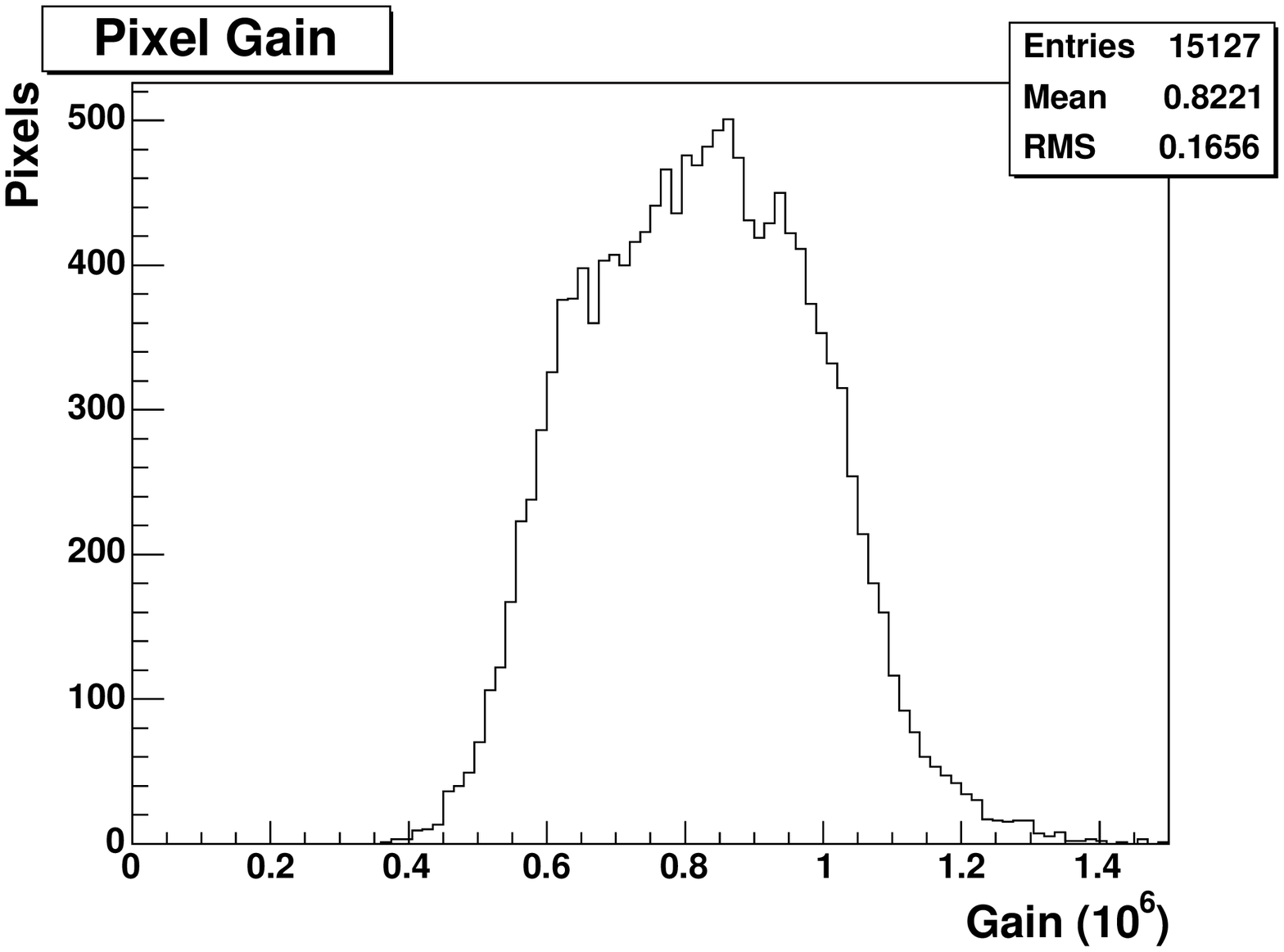}{Pixel gains}{One entry in this histogram
corresponds to one pixel for each of the 219 PMTs in the sample,
where each PMT was set to the operating high voltage.}
\end{figure}

\begin{figure}
\myfigcmd{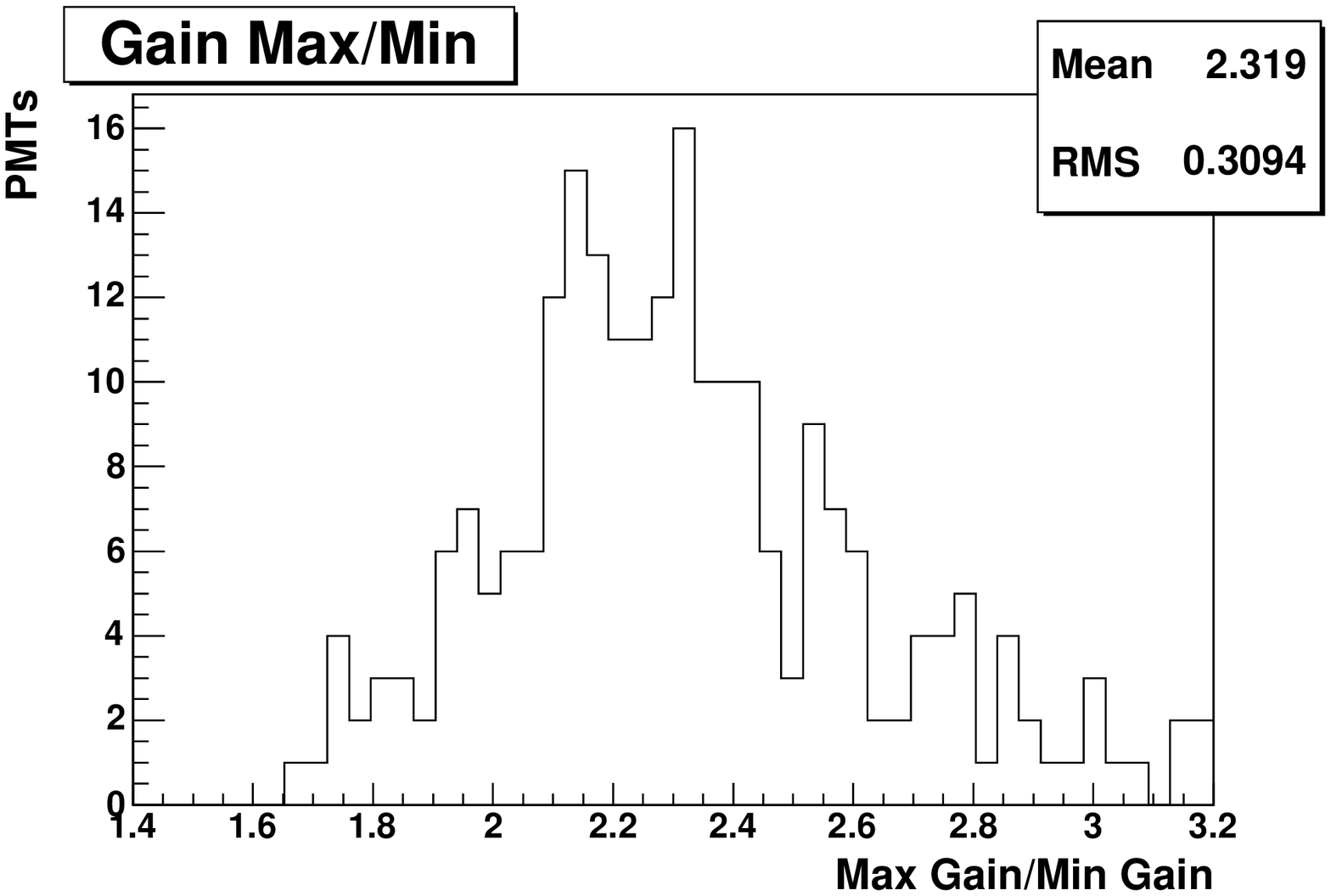}{PMT Gain Spread}{The ratio of maximum pixel gain
to minimum pixel gain is shown for all PMTs. }
\end{figure}

\begin{figure}
\myfigcmd{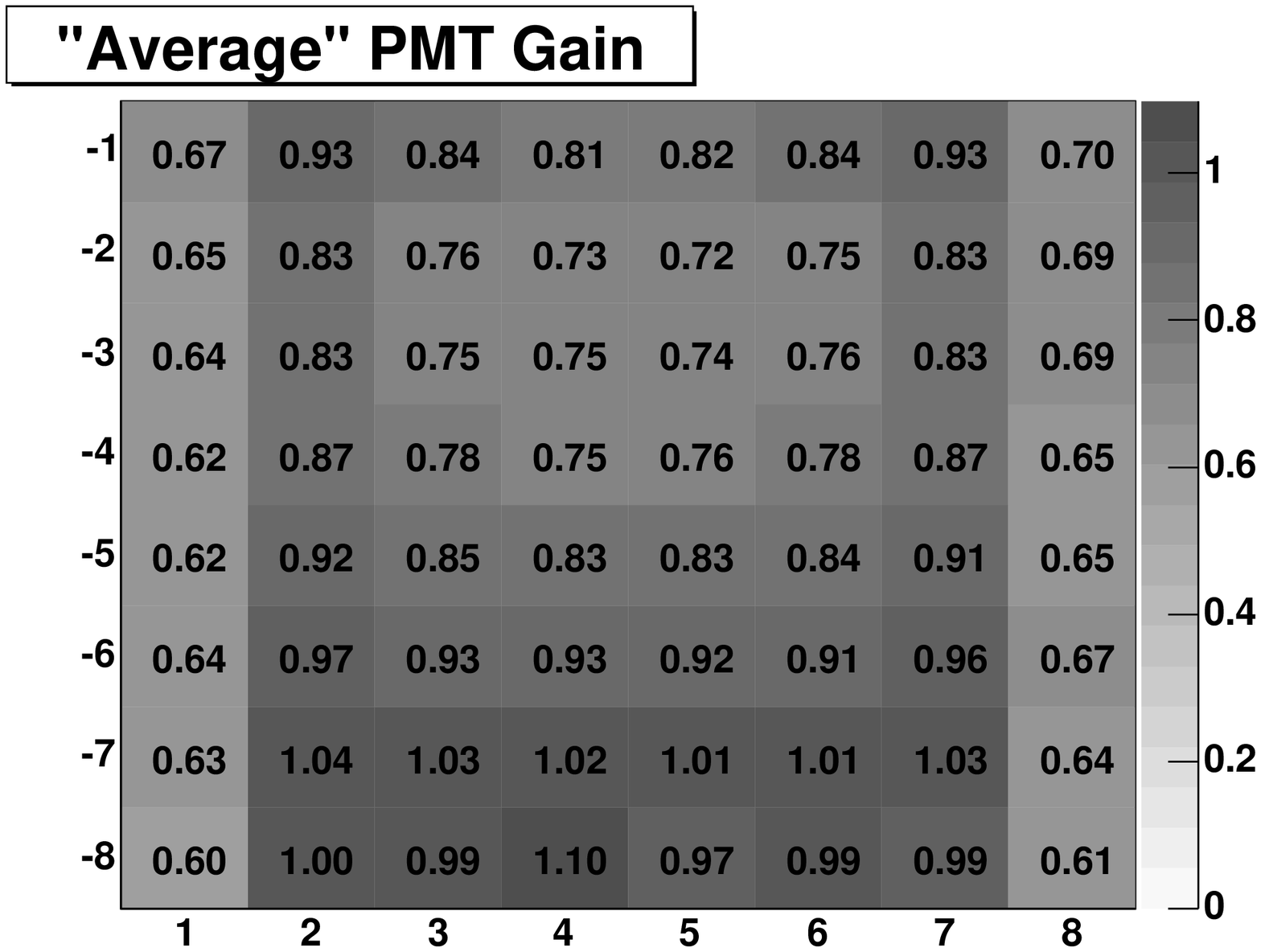}{Pixel Gain Pattern}{The average gain for each
pixel position is shown, averaged over 219 PMTs. Units are gain, to
be multiplied by $10^6$. Dynode slats run left to right.}
\end{figure}
 
Although no method was available to measure absolute quantum
efficiency of the PMTs, the number of measured photoelectrons
(for a given light level) could be compared between PMTs.  The
monitor PMT was used to correct for changes in the LED light level to
within one percent. The number of p.e. could then be used to compute
the ``effective efficiency'', meaning the product of quantum efficiency and
collection efficiency integrated over the spectrum of light shown in
Figure \ref{f:spectrum2}.  This effective efficiency was normalized to
the 520~nm quantum efficiency of three PMTs evaluated by Hamamatsu to
scale to an approximate absolute efficiency.

The efficiency of pixels on a given PMT was more uniform than gain,
typically within 10\% RMS over 64 pixels. Figure \ref{f:pixel_eff}
shows the relative efficiencies for all the pixels in the sample.  In
contrast to the gain measurement, there was no pattern in the
efficiencies of pixels at different positions.

\begin{figure}
\myfigcmd{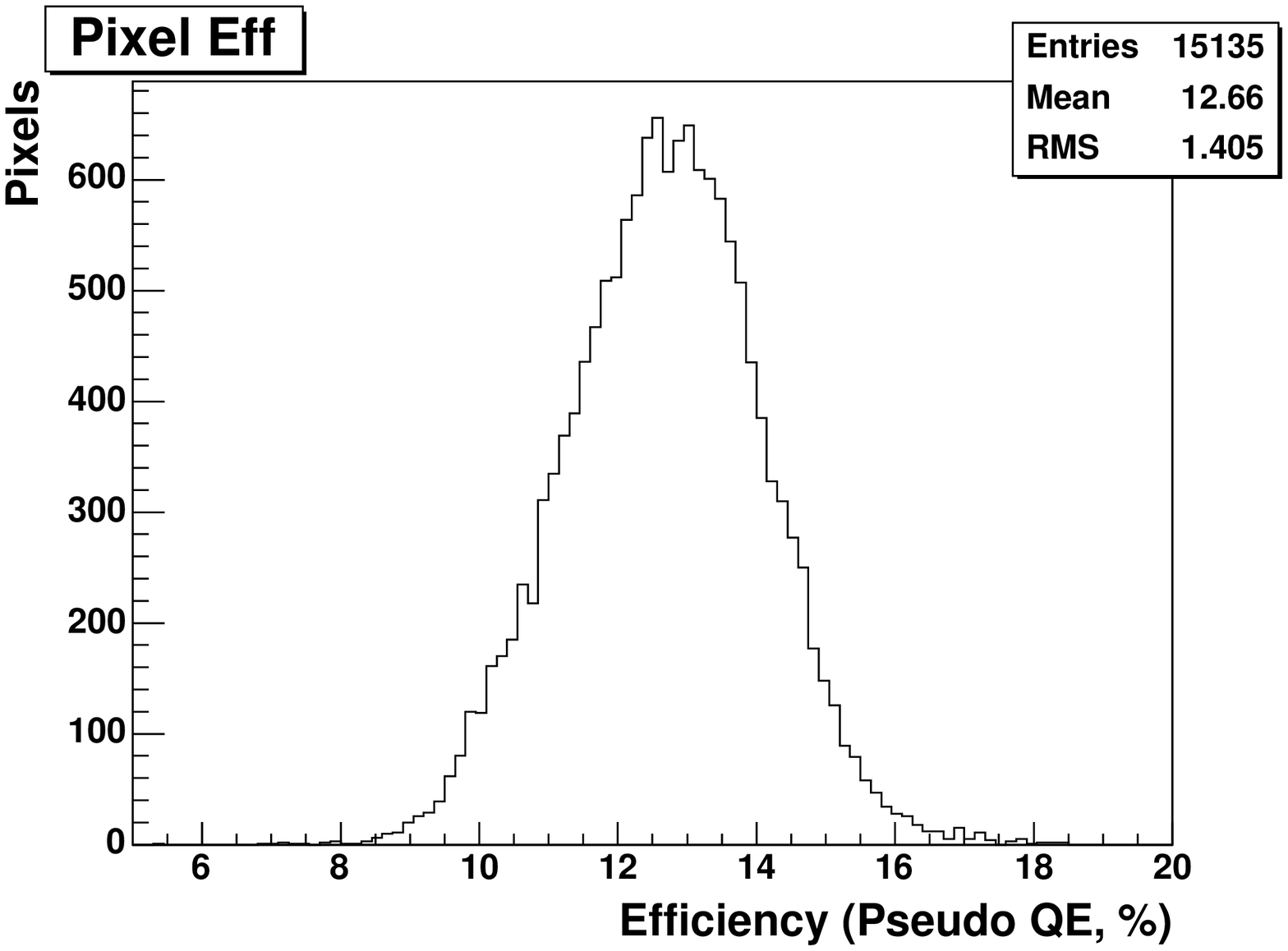}{Effective pixel efficiencies}{One entry in this histogram
corresponds to one pixel for each of the 219 PMTs in the sample,
where each PMT was set to the operating high voltage. The abscissa
is normalized to the quantum efficiency of three reference PMTs at
520~nm to give an approximate measure of absolute efficiency.}
\end{figure}

Six PMTs were rejected from the sample for having poor inter-pixel
uniformity.  One or more pixels on each of these PMTs had low gain
and sometimes low efficiency as well, creating an unacceptable overall
response. These pixels were frequently pixels 1--8 or 57--64, the
low-gain columns shown in Figure \ref{f:mean_pixel_gain}.

\section{Linearity} \label{s:lin}

The small size of the dynodes in an M64 led to a concern that
space-charge effects would be large enough to induce nonlinearity at
moderate light levels.  Figure \ref{f:log_linearity} shows the
nonlinearity of pulses in the region from $\sim$1000 to 70000 fC. The
true intensity of light (in p.e. per pulse) was calculated by assuming
the PMTs to be linear when illuminated with a filter to provide ~15~p.e, and
then calculating the incident light level for a different filter by
using the relative filter opacities.\footnote{Relative filter
opacities were measured {\it in situ} by illuminating the LED with DC
and measuring the green light with a photodiode.} The expected charge
for a given pixel was taken as the incident light times the gain times
the efficiency of the pixel.  

The magnitude of the nonlinearity varied greatly between pixels. The
size of the bars in Figure \ref{f:log_linearity} indicate the RMS
spread of pixels for the given charge (including a $\sim$1\%
statistical error).   This variance can be seen more
directly in Figure \ref{f:npe_at_nonlinearity}, which indicates the
illuminated light level at which pixels become nonlinear.  Some pixels
remain linear up to 350~p.e., while others become nonlinear at only 70-100~p.e.

For the purposes of MINOS, where signals are expected to be
approximately 5~p.e. (for muon tracks) to 100~p.e. (for dense electron
showers), these nonlinearities are acceptable. (An {\it in-situ}
measurement of the nonlinearity will be done in MINOS to ensure
accurate calorimetry.\cite{li-paper})

\begin{figure}
\myfigcmd{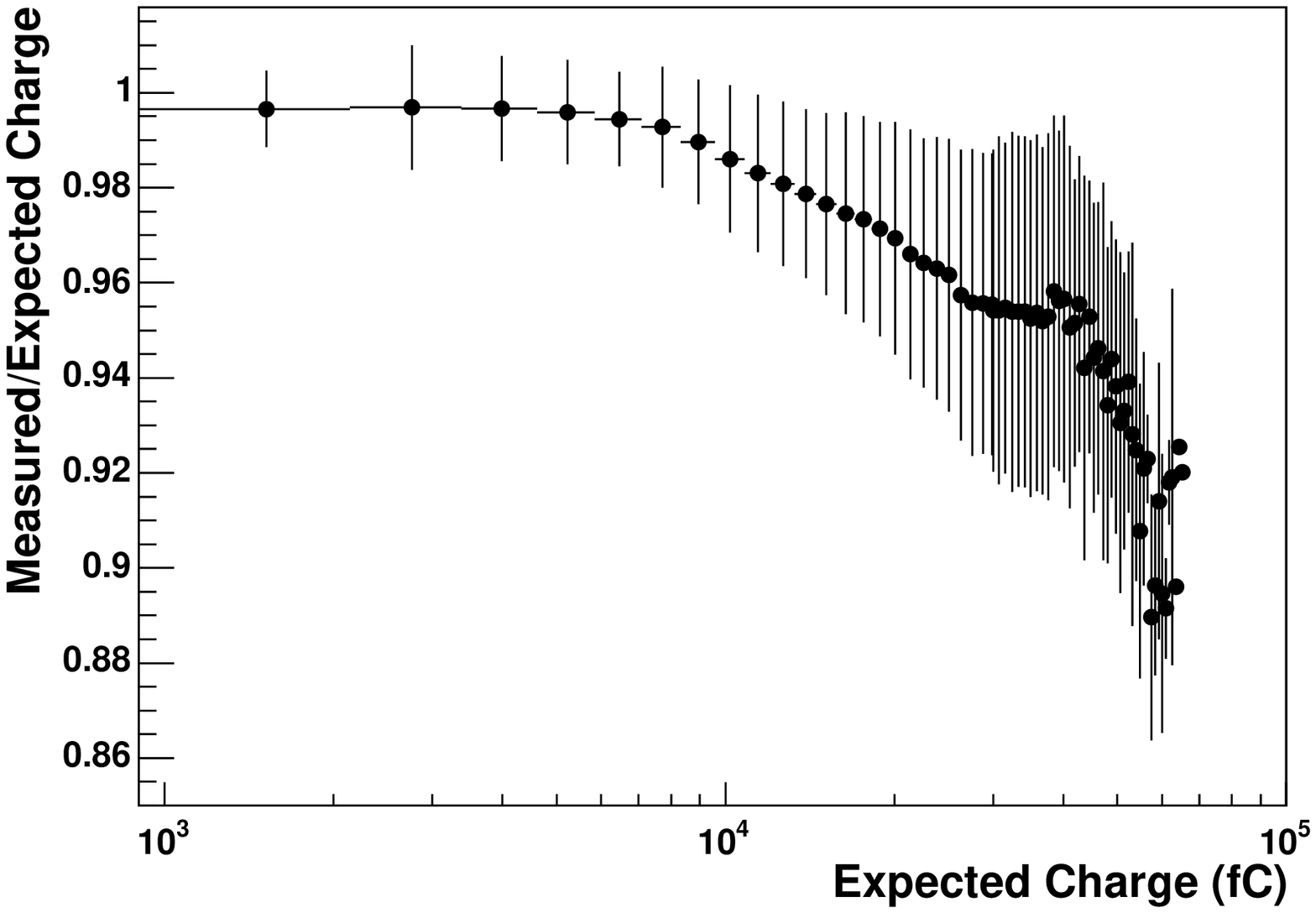}{Nonlinearity Curve}{The abscissa shows the
expected charge response for pulses at different light levels. (The
center of the plot is $10^4$~fC $\simeq$ 80~p.e.). The ordinate scale
shows the fractional deviation of the measured PMT charge from
linearity. Vertical bars represent the RMS variation amongst all the
pixels in the sample. The round markers show the average trend. Data
is shown for all pixels with light injected at seven different
intensities between $\sim$10 and $\sim$300 p.e.}
\end{figure}

\begin{figure}
\myfigcmd{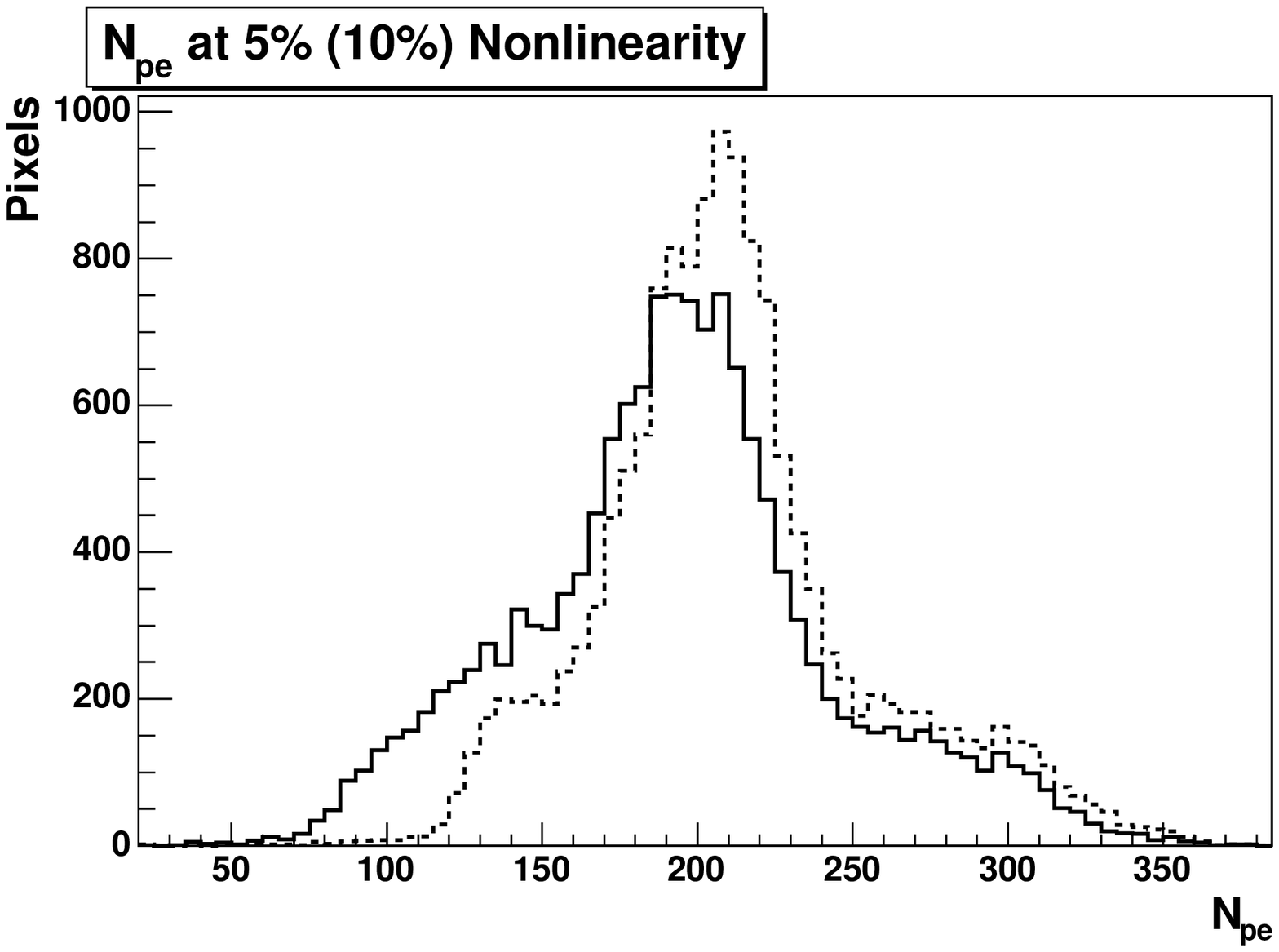}{Nonlinearity Thresholds}{The solid
(dashed) histogram indicates the values at which $N_{pe}$ illuminated
p.e. of light results in a charge that is suppressed by 5\% (10\%) below linear.}
\end{figure}

\section{Crosstalk} \label{s:xtalk}

Crosstalk was measured in the test stand by recording the integrated
charge on non-illuminated pixels while light was injected onto one
pixel.  Seven light levels between 10 and 200 p.e. were
used. 

The crosstalk within the readout electronics was small. The fraction
of charge leaked to non-injected pixels was measured to be less than
$6 \times 10^{-4}$ and typically $2 \times 10^{-4}$ between each pair
of pixels. This contribution accounts for only a small proportion of
the crosstalk observed.

Crosstalk between pixels on the PMT occurs by two different
mechanisms. The first form, ``optical'' crosstalk, is attributed to
primary photoelectrons getting multiplied in the wrong pixel's dynode
chain and thus giving a 1--p.e. signal in the wrong anode. The
second form, ``electrical'' crosstalk, is attributed to electrons
leaking from one dynode chain to another near the bottom of the chain,
resulting in a small fraction of the injected pixel's charge moving to
the wrong dynode channel or anode.
 
The non-illuminated pixels showed a small charge on every pulse that
was proportional to the charge seen in the injected pixel
(i.e. ``electrical'' crosstalk), and occasionally an additional large
charge consistent with a single photoelectron (i.e. ``optical''
crosstalk). The crosstalk was parameterized for every pair of pixels
for every PMT measured in the test stand.  Electrical crosstalk was
parameterized as the fraction of charge in the injected pixel that was
leaked to the non-injected pixel. This was found by observing the
shift in the pedestal peak of the non-injected pixel. Optical
crosstalk was parameterized as the fractional probability that a given
photoelectron would create a signal in the non-injected pixel.
Optical crosstalk was found by counting the number of single-p.e. hits
in the crosstalk pixel.  For both these mechanisms, the mean
charge seen in the crosstalk pixel was proportional to the charge seen
in the illuminated pixel.

A complete model was built by measuring crosstalk over different light
levels and different PMTs.  Both electrical crosstalk fraction and
optical crosstalk probability were constant with different intensities
of injected light. In general, the crosstalk averages were consistent
between different PMTs to within about 20\%. The values shown below
are taken as the average over all light levels.

Crosstalk was strongest between adjacent pixels, but was not limited
to this case.  Tables 2 and 3 show the fractions of charge that were
crosstalked by each of the two mechanisms for the nearest 8 pixels.
In summary, approximately 2\% of charge in the injected pixel was
leaked by electrical crosstalk, and approximately 4\% of the photons
incident on the injected pixel were detected on the wrong anode.
The precision of the values reflects only statistical errors, which
were small.  Systematic errors on the values in Tables 2 and 3 are
about 20\%. Poor accuracy was due to the difficulty in separating the
two small crosstalk signals. However, the total crosstalk (optical +
electrical) could be accurately measured as 6.9\% for the whole
sample.

In experiments where single-photon response is important the optical
crosstalk mechanism dominates. For instance, a signal of a single
photoelectron will appear to be on the incorrect pixel 4\% of the
time, leading to possible problems in interpretation of the data. For
large quantities of light the stochastic effects of the
optical crosstalk average out so that crosstalk is a simple
fraction of the incident light.

\begin{table}
\begin{center}
\label{t:elec-xtalk}
\begin{tabular}{rcccl}
\hline \hline
NW & & N & &  NE \\
 &0.0009 & 0.0036 & 0.0009& \\
W&0.0022 & -      & 0.0023&E \\
 &0.0010 & 0.0030 & 0.0010& \\
SW & & S & &  SE \\
\hline \hline
\multicolumn{4}{l}{Total to non-neighbors:} & 0.0063 \\
\multicolumn{4}{l}{Total to all pixels:}    & 0.0212 \\
\hline \hline
\end{tabular}
\mycaption{Electrical Crosstalk}{The values shown give the
average charge leakage fraction from an injected pixel to the eight
nearest neighbors, non-neighbors, and the total to all pixels. Dynode
slats run east-west.}
\end{center}
\end{table}

\begin{table}
\begin{center}
\begin{tabular}{rcccl}
\hline \hline
NW & & N & &  NE \\
 &0.0011 & 0.0054 & 0.0013 &   \\
W&0.0068 & -      & 0.0080 & E \\
 &0.0011 & 0.0056 & 0.0012 &   \\
SW & & S & &  SE               \\
\hline \hline
\multicolumn{4}{l}{Total to non-neighbors:} & 0.0069 \\
\multicolumn{4}{l}{Total to all pixels:}    & 0.0374 \\
\hline \hline
\end{tabular}
\label{t:opt-xtalk}
\mycaption{Optical Crosstalk}{The values shown give the average 
probability for a PE from an injected pixel to crosstalk to each the
eight nearest neighbors, non-neighbors, and any other pixel.}
\end{center}
\end{table}

\section{Dark Noise} \label{s:dark}

M64s were tested for dark noise by taking data with no light on the
PMT, with a high voltage of 950~V. The readout system was pulsed
$4 \times 10^7$ times, for an integrated charge-collection time of 44
seconds.  Rate was determined by counting the number of readouts for which
the integrated charge was greater than a threshold, defined as
one-third of a p.e. for the lowest-gain pixel on that PMT.  Before
conducting the dark noise measurement, the PMTs were under high
voltage and exposed to no light for a minimum of 12 hours. The dark
noise measurement itself was conducted over two days.  The ambient
temperature was controlled to be at $20 \pm 2$~C.

MINOS specified that PMTs should have total anode noise rates (for all
64 pixels) of less than 2~kHz, but the rates measured for most PMTs
were far lower.  The average noise rate was 260 Hz per
PMT. Approximately 10\% of the PMTs had noise rates greater than 500
Hz; approximately 5\% had rates greater than 1000 Hz.  Two PMTs were
rejected from the sample for very high noise rates. The spectra of
noise pulses was consistent with a single photoelectron spectrum. It
was frequently found that a single pixel would be considerably more
noisy than all the others on a PMT; the noisiest pixel contributed on
average approximately one third of the total dark noise.

In MINOS, the PMT readout is triggered by a discriminator connected to
the tap on the 12th dynode. The signal from this dynode is similar to
an analog sum of the individual pixel signals. The rate of pulses on
this dynode, using a threshold equivalent to the $1/3$ of a p.e. on
the lowest-gain pixel, was found to give consistent results with the
method described above.

\section{Conclusions}

We have found that M64s may be used to provide good measurement of
light from wavelength-shifting optical fibres for intensities of
1--100 p.e., the measurement of interest to MINOS.  The variance in
gain between pixels on a PMT is 25\% RMS. Quantum efficiency is
similar between PMTs.  Excepting a handful of rejected PMTs, the
single-p.e. peak was well-resolved from the pedestal at our operating
voltage.  The PMTs are typically linear for pulses less than 100~p.e. at $8
\times 10^5$ gain (i.e. 13~000~fC), and are comparable to the
16-pixel R5900-00-M16 PMTs\cite{Lang:2001tx}. The M64s have low dark
noise, typically 4~Hz per pixel, due to their small pixel area.

Crosstalk in M64s can be a significant problem, particularly if they
are used to count single photons since photoelectrons can get
collected by the wrong dynode. However, if typical signals are larger,
and pixel occupancy small, the adjacent crosstalk signals are easily
identified.

In each of the studies described above (single-p.e. response,
uniformity, linearity, crosstalk, and dark noise) no significant
variation was seen between different delivery batches, or between the
R5900 and R7600 models.

We have found M64 PMTs to meet or surpass requirements for
reading out scintillator with wavelength-shifting fibres. A total of
about 5\% of PMTs were rejected for not meeting these requirements.
The MINOS near detector, under construction at the time of this
writing, will be employing these devices on a large scale.

\section{Acknowledgements}

We gratefully acknowledge the help and assistance provided by
Hamamatsu Photonics Ltd, and thank them for allowing us to reproduce
some of their specifications here.  We appreciate the assistance of
B. Brooks for his help assembling our LED pulser, and P. Sullivan for
his work designing and producing the PMT bases.  Most importantly, we
would like to thank the members of the MINOS collaboration, in
particular G. Drake, for his help with instrumentation, K. Ruddick for
his work on WLS spectra, J. Alner and T. Durkin for work on PMT boxes
and M. Kordosky, and P. Vahle for their work in early pioneering of
the techniques with M16 PMTs. This work was funded by the Particle
Physics and Astronomy Research Council to which we are grateful for
support.

\nocite{hamamatsu_book}
\nocite{early_paper}

\bibliographystyle{elsart-num.bst}
\bibliography{m64paper}

\end{document}